# PREDICTING SPLASH OF A DROP IMPACTING A THIN LIQUID FILM

**S. Rajendran, M. A. Jog and R. M. Manglik**

Thermal-Fluids and Thermal Processing Laboratory,

Department of Mechanical and Materials Engineering,

University of Cincinnati, Cincinnati OH 45221

**ABSTRACT**

In most spray coating and deposition applications, the target surface may be initially dry but with continuous drop impact a thin layer of liquid film is formed on which further impingement occurs. An experimental study of the process of drop impact on a thin stagnant film of the same liquid and the subsequent drop-film interactions is carried out. The impacting drop results in either liquid deposition or it can cause prompt or delayed splash. Deposition occurs when the drop merges with the liquid film without generating secondary drops. Splash results in the production of secondary drops either at the instant of impact (prompt splash) or through a delayed break-up of the rim of the crown formed as a result of the impact (delayed splash). Experiments are conducted to characterize the phenomena using five different Newtonian liquids and by varying drop impact diameter and velocity. The liquids are chosen so as to cover a wide range of liquid properties (viscosity and surface tension). A high-speed digital camera Hi-D cam – II version 3.0 – (NAC Image technology) is used to capture the drop impact dynamics. The threshold of splashing is found to be related to drop size, impact velocity, liquid properties and thin film thickness.

Experimental analysis of the significance of inertial, viscous and capillary forces in determining the splash/no-splash (or deposition) boundary helps in establishing an empirical correlation for the same. The splash/no-splash outcomes predicted by the proposed correlation agree well with experimental data available in the literature.

**NOMENCLATURE**

<u>Parameters</u>

| | |
|---|---|
| $g$ | Gravitational acceleration [m/s²] |
| $v$ | Velocity of the issuing jet [m/s] |
| $d$ | Impacting drop diameter [m] |
| $h$ | Thin film thickness [m] |

<u>Greek Symbols</u>

| | |
|---|---|
| $\rho$ | Density of the liquid [kg/m³] |
| $\mu$ | Dynamic viscosity of the liquid [Pa.s] |
| $\sigma$ | Surface tension of the liquid [N/m] |

<u>Non-Dimensional Numbers</u>

| | | |
|---|---|---|
| $We$ | Weber Number | $\left(= \rho v^2 d / \sigma\right)$ |
| $Re$ | Reynolds Number | $\left(= \rho v d / \mu\right)$ |
| $Oh$ | Ohnesorge Number | $\left(= \mu / \sqrt{\rho \sigma d}\right)$ |
| $Mo$ | Morton's Number | $\left(= g\mu^4 / \rho\sigma^3\right)$ |
| $H^*$ | Dimensionless film height | $\left(= h/d\right)$ |

# 1 INTRODUCTION

Improving our understanding of drop-liquid layer interactions during drop impact is useful to a multitude of engineering applications ranging from coatings, spray paintings, ink-jet printing, liquid atomization, cleaning, and metal annealing, where, the phenomena is commonly encountered. Drop impingement is also a key mass transport in soil erosion, spores and microorganism dispersion, and pesticide coatings. The working liquid used in these applications are chemicals that could get air-born and pollute the immediate environment if the impingement resulted in a splash. Apart from the environment, these could also lead to chemical and/or biological occupational risks for the workers, leading to dermatological and respiratory ailments. Industries involving chemical sprays often produce drops that contain surfactants, biocides, chelating agents and, fatty acids, all of which can be harmful for the workers. These have been noted to increased occurrence of asthma, hypersensitive pneumonitis and cancer of esophagus, stomach, larynx etc.[2]. In order to minimize catastrophic circumstances resulting from outcomes of impingements, Occupational Safety and Health administration (OSHA) and National Institute of Occupational Safety and Health (NIOSH) have defined permissible exposure limits that the industry is required to meet [3]. Resolving and understanding the phenomena of drop impact would assist in exercising better control of the impact and can have various pay-offs from enhancing industrial needs to lowering needless worker exposures. A means to predict and control the splash occurring while maintaining the functionality of the process, becomes crucial.

As the phenomenon of drop impact is inherently complex, its physical understanding is far from complete. Worthington [1] was perhaps the first to document the resultant patterns created by falling drops of water and mercury on a dry surface in 1876. Since then the phenomena of drop impact remains the focus of research due its practical relevance and innate complexity. Improvements in experimental techniques and high-speed imaging in the recent past have greatly helped propel advancement in the understanding of the drop impact process [XX]. These, coupled with advancements in computational power and numerical techniques has helped obtain distinct features of drop impact (formation, propagation, crown growth, splashing limits, etc.) and the role of drop characteristics such as size, shape, and velocity [XX]. Except for a few applications, most spray impacts occur on a thin film accumulated after the impact from previous droplets. The interaction of a liquid drop impinging on a layer of thin film is addressed in the present work.

An extensive review of drop impact dynamics was provided by Yarin [4] who showcases the dependency of impact dynamics on a number of parameters: drop size and impact velocity, drop and film liquid properties (density, viscosity, rheology), interfacial tension, surface properties (roughness, contact angle, wettability), and thermal properties. When a drop impinges on a thin liquid film, it leads to different regimes of liquid movement that can broadly be classified under *deposition* or *splashing*. Illustrations of these regimes are provided in Fig. 1. *Deposition*, post impact of a drop, involves the drop merging with the liquid film without forming secondary drops. *Deposition* may or may not include capillary surface waves that form a *crown*, before merging with the liquid film. *Splashing* happens when secondary drops are discharged as a drop impacts the thin film. At certain conditions on impact, an ejected jet is formed at the neck (small region between the drop and the thin film). The ejected jet spreads out and tilts upwards, forming a *crown* whose rims are unstable and break-up into secondary drops before merging into the thin film. This was termed *crown splashing* by Deegan et al. [5] and by Josserand and Zaleski [6]. *Prompt splash* is a supplementary phenomenon that could occur when secondary drops are formed at the instant of drop impact on the thin film. Deegan et al. [5] studied the complexities of splashing and found that during a prompt splash, the ejected jet shot out from underneath the drop parallel to the fluid layer. Their study delineated the different regimes of drop impact based on two parameters: Weber number (We) and Reynolds number (Re). They present a correlation to demarcate occurrence of prompt splash from crown splash. *Delayed* or *crown splash*, occurs when the liquid crown breaks up at or beyond the expansion of the crown. Both experimental and analytical methods have been used to find the splashing threshold. While some of the previously highlighted work here use analytical means to identify a threshold limit, Yarin and Weiss [7] experimentally defined a velocity threshold for drop splashing that is based on the frequency of falling drops, the viscosity, surface tension and diameter of the impacting liquid drop.

Three dimensionless parameters have been commonly used in scaling drop impact and splash threshold: Weber number $\left(We = \frac{\rho v^2 d}{\sigma}\right)$, Ohnesorge number $\left(Oh = \frac{\mu}{\sqrt{\rho \sigma d}}\right)$ and dimensional thin film height $\left(H^* = \frac{H}{d}\right)$. Cossali et al.[8] tested the splashing threshold by analyzing drop impact experiments that involved a wide range of liquids on different target surface conditions. The H* in these experiments were maintained between 0 and 1. They proposed a correlation to estimate splashing limit; $We\ Oh^{-0.4}$ = K = constant, where K depends on the non-dimensional film thickness to predict the deposition-splashing limit. The value of K was found to increase with increasing H*. This parameter K was initially developed to describe drop impact on dry solid surfaces [7, 9] and has been adapted by some researchers to describe drop impact on thin films. Multiple viscous liquid impact on thin films was tested by Rioboo et al.[10] to define two thresholds of impact outcomes: the deposition-crown and crown-splash limits. Both these thresholds were seen to be influenced by the film height H* when the film was very thin (H* < 0.06) but were seen to be a constant at larger film thicknesses (0.06 < H* < 0.15). For each of these thresholds, values of K at different film thicknesses have been defined in their study. Okawa et al. [11] find a similar relation in their experimental study for the two thresholds as defined by Rioboo et al. [10], though their value of K for the two thresholds are different. Although their film height ranges were higher (H* = 0.48 – 68) than the other studies discussed here (with H* < 0.1), their findings for the onset of the different regimes of drop impact are very similar to that of Rioboo et al. [10]. In addition to defining the onset of the regimes of drop impact dynamics, they also looks at secondary drop characteristics and its dependence on the K parameter. The difficulty in producing and maintaining small film thicknesses motivated Wang and Chen [12] to explore reliable and repeatable means to produce thin films of a surface. With the help of their new technique of maintaining thin films with higher confidence, they found that the critical splash level was insensitive to film height when the film is sufficiently thin (H* ≤ 0.1). Different concentrations of glycerol water mixtures were used to experimentally find splashing criterion. They expressed the criterion as We = constant for a given liquid and different values of the constant with changing viscosity of the test liquid. Similar observations were made by Vander Wal et al. [13] where Weber number was found to be the determining parameter to establish the onset of splash for thin films (H* ~ 0.1). However, unlike the previous study, they find that the splash criterion is not dependent on the viscosity of the liquid. Experimental examinations using multiple viscous liquids were used to understand the onset of splashing both on dry surfaces and on thin liquid films. Their work also highlights the effect of surface roughness when a drop impacts a dry surface.

Apart from experimental studies, several numerical techniques have helped improve our understanding of the onset of splashing. Based on the concept of kinematic discontinuity proposed by Yarin and Weiss [7] a model for the propagation of the crown formed during drop impact was proposed by Trujillo et al. [14]. Their solution resulted in defining a time scale beyond which it was found that the crown characteristics were independent of upstream conditions. The model also attempted to resolve the effect of surface film characteristics and was validated with experimental data. A 2D flow model was numerically simulated using VoF method by Coppola et al. [15]. Their work concentrated on validating the proposed kinetic discontinuity theory numerically.

In this current work, the onset of delayed splash is studied experimentally. While there are a few studies that outline the onset of splashing, there is variability in their predictions based on their proposed correlations. In this study, we attempt to understand the underlying reason for this disparity. Many studies have concentrated on using water as the working liquid and the proposed correlations do not truly take into account the significance of viscosity of the liquid. In this experimental study, five different liquids with varying liquid properties are tested with the aim to create an improved understanding of the phenomena of drop impact on Newtonian liquid films.

## 2 EXPERIMENTAL SETUP

Consequence of a single drop impingement is studied experimentally, as it impacts a thin film formed from prior impinging drops. The thickness of this thin film formed prior to impact was maintained at H*~ 0.1. The film thickness is maintained and is left undisturbed prior to the experimental observation. Fig. 2 provides a sketch of the experimental apparatus used. A container made of acrylic sheets was used as the main observation chamber and functioned as the reservoir/collecting tank. The container had a raised acrylic platform at the center, where the copper target surface was installed. The walls of the container were made to be large enough to avoid interference with the splashing droplets and to provide for clear viewing of the ensuing post-impact dynamics. Impacting drops were generated at a steady rate from a NEXUS 3000 syringe pump. The change in drop release height along with adjustments to the flow rate, provided a range of drop impact velocities. The flow rate was varied from 0.5 ml/min to 8 ml/min for the different cases tested experimentally. This resulted in a range of velocities from 1 m/s to 3 m/s for drop diameters ranging from 3.5 mm to 5.2 mm. The drops were released from a prescribed height above the target surface, through a stainless steel circular needle. The needle was always maintained perpendicular to the target surface using a bubble level. The target surface has no inclination and is maintained parallel to the ground. Since the target surface was much larger than the impacting drop, the increase in film thickness with each impacting drop was found to be negligible. As has been discussed in Fig. 1, when a single drop impacts the thin liquid film, it may either result in a splash or in deposition. Varying drop size and velocity of impact helped produce a range of outcomes to that are analyzed further.

To study the phenomena of drop impact, a high-speed digital camera (Hi-D cam – II version 3.0 – NAC Image Technology) was used. The camera system was kept at an appropriate angle to view the impact without causing hindrance to the phenomena. Images were recorded at a shutter speed of 1/2000 s and a frame rate of 500 fps with the camera placed at about 1.5 ft. from the target surface. A single bulb focusing light system (ARRI) with glossy aluminum reflectors was used to obtain clear and high contrast images. The light system was focused on a white screen that was placed parallel to the plane of viewing. Care is taken to ensure that the light system is kept at some distance from the test area to avoid heating the immediate environment of the apparatus. The high-quality images thus obtained were analyzed using the image processing software, Image-Pro 4.0 (Media Cybernetics). The impacting diameter of the drop was calculated based on the area occupied by the drop, assuming the drop to be spherical and therefore, a circle in the image plane. Drop impact velocity was determined by comparing the position of drops, prior to impact, in successive frames with respect to a fixed point in the frame. The impact velocity was not a constant due to gravitational acceleration and drag force acting on the drop. Therefore, only three appropriate images before impact were considered for calculations. For each set of experiment, images were calibrated using the outer diameter of the stainless-steel needle. Each experiment was repeated to check for repeatability and the average measurements are used for analysis.

To help understand the influence of liquid properties on the splashing phenomena, different liquids with varying properties were used for the experiment. Table 1 presents the properties of the five liquids used. Mixtures of water and propylene glycol by volume were used as working fluids to obtain a range of liquid properties for the experiment. The values of viscosity and surface tension of these mixtures were established by using the AR 2000 controlled stress/controlled shear rate Rheometer and the SensaDyne QC6000 Tensiometer respectively.

## 3    RESULTS AND DISCUSSION

Controlled experiments, as described in the previous section, are performed to understand the behavior of drops impinging on a thin film of the same liquid. High-speed images obtained during the experiment are used to establish the outcome of each impact (as deposition or splash). Past studies on thin film drop impact have suggested correlations based on their data to find a means to bound the outcomes of the impact and to predict the onset of splashing. These are compared with past experimental data from other research studies along with data from this current study and are presented in Fig. 3. In Fig. 3, the data points for splash and non-splash (i.e. deposition) events are shown in blue and red colors, respectively, at the corresponding Weber and Reynolds numbers at impact. The suggested correlations for determining the onset of splashing are primarily of two forms. While Cossali et al. [8] and Rioboo et al. [10] have correlations of the form of $WeOh^{-0.4} = We^{0.8} Re^{0.4} = C_1$, Vander Wal et al. [13] and Wang and Chen [12] propose that $We = C_2$ provides the demarcating line beyond which splashing would occur. In the former type of correlation, both studies propose different values for the constant $C_1$. Rioboo et al. [10] observe $C_1$ to be a fixed value for thicker films ($0.06 < H^* < 0.15$) and to be a function of the film thickness for very thin films ($H^* < 0.06$), while Cossali et al. [8] observed $C_1$ to always be a function of the film thickness. In the second relation ($We = C_2$), while Vander Wal et al. [13] identifies $C_2$ to be invariant of the liquid property, Wang and Chen [12] note that the constant value of We is dependent on the Oh number of the liquid. Thus, the constant We line in Wang and Chen's [12] findings is parallel to Vander Wal et al.'s [13] and is different for each liquid shown in the plot (Fig. 3). The plot shows the performance of these correlations with experimental data from the current study as well as those in past literature for $H^* \sim 0.1$. While it is noted that most splashing events occur beyond the relation We $Oh^{-0.4}$ = $We^{0.8}$ $Re^{0.4}$ = $C_1$, this is an over-prediction of the onset of splashing as certain deposition events too are observed beyond this curve. The We = $C_2$ on the other hand, is seen to heavily under-predict the splashing threshold for certain liquids while over-predicting the same for others. If $C_2$ were to depend on liquid property and change for each liquid as suggested by Wang and Chen [12], overlapping data points for splash and deposition should not be observed. In the current study we have obtained closely spaced data points to determine to exact onset of splashing for the liquids used. In the current plot (Fig. 3), certain splash and deposition points are noted to overlap for these closely spaced data points. The primary form of these correlations are in terms of Oh and Re. To discern the performance of this relation, our experimental observation for two liquids: water and ethylene glycol, are plotted along with the predictive curves for correlations discussed before (Fig. 4). As before, the Vander Wal et al.[13] and Wang and Chen [12] equations are parallel to each other as are the Rioboo et al. [10] and Cossali et al. [8] curves. It is interesting to note that the slopes of these two sets are similar on the Oh vs. Re plot.

While comparing the data for water (Fig. 4a) it is conceivable to see that for smaller drop diameters (< 4.5 mm), modifying the intercept of the two established curves can demarcate the splash and deposition limits. However, for larger drop diameters (dotted red oval in Fig. 4a), the trend seems to deviate. Fig. 4b shows the behavior of Ethylene glycol splash and deposition data points in relation to the available correlations. There is an overlap between the splash and deposition events showing that this scaling would not be appropriate to demarcate the current experimental data. These observations indicate that considering only Re and We (and/or Oh = $We^{0.5}Re^{-1}$) as the governing dimensionless parameters is insufficient and indicates existence of some other important forces that may be significant in governing the drop impact outcome.

To identify all the relevant significant parameters, the effect of different working liquid properties is considered in this current experimental study. The analysis ensues from the examination of high-speed images captured during drop impact on a thin film in stagnant ambiance. The impact of water drops on a thin water film is shown in Fig. 5. The occurrence of four different phenomena, that can be broadly categorized into splash and deposition, are presented here. Fig. 5 a and b show splashing occurring, while Fig. 5 c and d show deposition. Images are captured at an interval of 0.002 s. As the drop impacts on the thin film surface, at the instant of impact, thin ejected jets emitting secondary drops produce a splash. This is prompt splash and its progression is shown in Fig. 5 a. The ejected jet formed at impact resembles a crown as it continues to expand. The capillary surface waves formed on the crown create fingers of the crown that extend upwards giving it the distinct structure. These surface waves could then grow into fingers that could then break-up or pinch-off to form the secondary drops. In Fig. 5 a, while the drops formed from the prompt splash are seen at the instant of breakup at 0.002 s, the fingers of the crown are seen to grow and ultimately break up into drops at 0.14 s while falling back into the thin film layer. Fig. 5 b shows the case of a delayed or crown splash, where, at impact, no prompt splash occurs. However, at impact, the formation of capillary waves can be seen on the surface of the crown. These propagate to form the fingers of the crown that breakup at 0.012 s. During crown splash, the droplets formed are noticed to be much larger and uniform in comparison to prompt splash in Fig. 5a. Fig. 5 c and d present deposition of the drops. In Fig. 5c the capillary surface wave formed at impact at 0.002 s is seen to propagate and form fingers of the crown. These fingers, however, do not further breakup into drops before deposition on the thin liquid film. The drop in Fig. 5d deposits on the thin film layer without forming any capillary waves on impact (0.002 s). Minor surface undulations are observed at 0.01 s. These undulations, however, do not form finger-like structures before deposition. Accordingly, there can be two types of deposition: fingered and non-fingered. By inspecting Fig. 5 c with respect to a and b, it is noted that both an increase in diameter and an increase in impact velocity affects droplet splashing on a thin film. The diameter of the resulting crown on the thin film is seen be a function of the drop diameter. So, for the same velocity of impact, a smaller drop results in a smaller crown diameter with lower splashing (Fig. 5 a and c) and unquestionably, for the same drop diameter, a larger velocity results in splashing (Fig. 5 b and c).

To understand the influence of different forces on the phenomena of splashing, the effect of liquid properties is studied in addition to the drop size and drop velocity. As the drop impacts the thin film, it is postulated that the base of the crown has a velocity comparable to the impact velocity. This velocity of growth of the crown base, decreases with crown development until the crown finally collapses on the thin film. Therefore, it is hypothesized that, the spread of the impacting drop primarily receives resistance from the thin film. The decrease in velocity of the crown base can therefore be related to the resistance offered by the liquid. This resistance points to the viscous property of the liquid. Thus, it becomes significant to understand the role of the viscosity of the liquid. Time progression of two different liquids (EG and 50% PG) drops, impacting a thin film with similar velocities are shown in Fig. 6. The Weber number of these two cases shown in Fig. 6 is nearly the same. For the EG drop, at 0.004 s, the rim of the crown is still uniform and no undulations are observed. Instability on the crown rim is not observed until deposition occurs at 0.018 s. For 50% PG, though the drop diameter is smaller than that of EG in Fig. 6a, minor surface waves are noticed at 0.006 s. These disturbances grow and ultimately result in a modest splash at 0.018 s. Though the diameter of the EG drop is higher than the 50% PG drop, the EG drop does not splash. As the viscosity of EG is higher than 50%PG, EG liquid film would offer more resistance to the growing crown and could suppress splashing by a greater measure in comparison to 50% PG. Consequently, it can be concluded that viscosity of the liquid affects the spread of the liquid crown layer thereby splashing.

Fig. 7 illustrates two liquid drops, of 25% PG and of 75% PG, impacting a thin liquid film of the same liquid. The two cases considered have similar velocities but different drop diameter, Re and We. For 25% PG, at 0.002 s, the impact is seen to generate capillary waves. The formation of capillary waves is not observed on impact for 75% PG. For both the liquids shown, the capillary waves lead to the formation of fingers of the crown that eventually break-up to form droplets. For 25% PG (Fig. 7a), this breakup occurs more easily than for the 75% PG drop. Also, the number of splash drops (secondary drops formed after splashing) is more for 25% PG. As the liquid crown is spreading, the instability on the rim leads to crown formation and possible breakup leading to the formation of droplets. In addition to this instability driven crown breakup, the crown edges are constantly looking to minimize their surface energy. This surface energy is governed by the surface tension of the liquid and is lower for a unit volume of a drop than for the same unit volume of the crown. As seen in Table 2, the surface tension of 25% PG is higher than that of 75% PG. As surface tension governs the surface energy, the 25% PG drop splashes more violently than the 75% PG. It can thus be concluded that a higher surface tension helps in breaking up the crown rim hence leading to formation of secondary droplets.

Based on the analysis of experimental observations in Fig 5, 6, and 7, it is hypothesized that while the crown rim formation and its subsequent growth is resisted by the viscous force of the liquid in the stagnant thin film, the breakup of the crown rim into secondary drops is dictated by the surface tension force acting to minimize surface energy at the crown rim edge. These two interactions can be expressed in terms of dimensionless parameters of Re and Bo ($Bo = \frac{\rho g d^2}{\sigma}$), where Re denotes the inertial to viscous force interaction and Bo denotes the gravitational to capillary (surface tension) force interplay. The influence of these two parameters on defining splashing threshold is further explored. A plot of Re vs. Bo with the current experimental data is shown in Fig. 8a. Further analysis provides the following relation between Re and Bo to predict the onset of splash:

$$Re = C_3 Bo^{0.1} \quad (1)$$

It is observed that the demarcating curve is dependent on the property of the liquid in consideration. Further experimental evaluation shows that $C_3$ is a function of Morton's number $\left(Mo = \frac{g\mu^4}{\rho\sigma^3}\right)$ and is given by: $C_3 = 60.95 \, Mo^{0.2}$. Thus, the semi-empirical relation to predict the onset of splash of drops in thin films can be expressed as in equation (2).

$$Re = 60.95 \, Mo^{-0.2} Bo^{0.1} \quad (2)$$

The curves shown in Fig. 8 a and b correspond to the above identified relation. To examine the validity and accuracy of this relation, experimental data from past studies that were used in Fig. 3 are tested against this proposed new relation in equation (2). The plot in Fig. 8b shows that the current expression is able to predict the onset of splash with good accuracy for all the experimental data assessed from past literature. It must be noted that, as this relation is obtained for H* ~ 0.1, the validity of this expression is limited to impact on thin films with dimensionless heights (H*) ~ 0.1.

It is to be noted that in this examination, the effect of gravitational forces in not neglected unlike some post research studies. Cossali et al.[8] thought that post-impact, the crown propagation was driven by the opposing surface and inertial forces while viscous forces dampened the propagation. Gravity was thought to act in a manner similar to the surface forces. Bond number (Bo) represents the relative importance of gravitational force to surface force. Since Cossali et al.[8] looked at situations with Bo ≤ 1, they reason that gravity could be neglected. In the current study, the Bo for most cases is ≥ 1 and hence the effect of gravity is not negligible. The derived onset of splash expression (Equation 2) is seen to hold true for both cases.

## 4 CONCLUSIONS

An experimental investigation of drop impact dynamics on thin liquid films is carried out. Multiple Newtonian liquids are used to cover a range of liquid properties and to understand their effect on the drop-liquid film interactions post impact. In addition, a range of drop diameters and impact velocities are considered while maintaining dimensionless film height (H*) of 0.1. The detailed analysis of high-speed images utilized to capture rapid developments of drop impact have been discussed. The effectiveness of the predictive correlations presented in past studies for determining the onset of splashing have been compared with the past and current experimental data (all comparisons are for H* ~ 0.1). These correlations are found to have some inconsistencies in their predictions, thus indicating the presence of potential significant considerations that have previously been overlooked. The experimental analysis of the phenomenon under consideration signifies the importance of the liquid properties, specifically viscosity and surface tension, apart from contributions of the drop diameter and impact velocity. These observations lead to two important hypotheses: (1) drop impacting a thin film encounters viscous resistance from the thin film as the drop spreads along the film forming a crown, and, (2) surface tension of the liquid in the crown rim aims to minimize the surface energy of the rim thus governing its breakup and the production of secondary drops. Consequently, this leads to two governing dimensionless parameters: Re and Bo respectively. Based on these two parameters, a predictive semi-empirical correlation is developed for demarcating the onset of splash when a Newtonian drop impacts a thin liquid film. In addition to the current experimental data, the predictive correlation agrees well with data from past experimental studies.

## 5    ACKNOWLEDGEMENT

This research study was partially supported by the National Institute for Occupational Safety and Health Pilot Research Project Training Program of the University of Cincinnati Education and Research Center Grant #T42/OH008432-08.

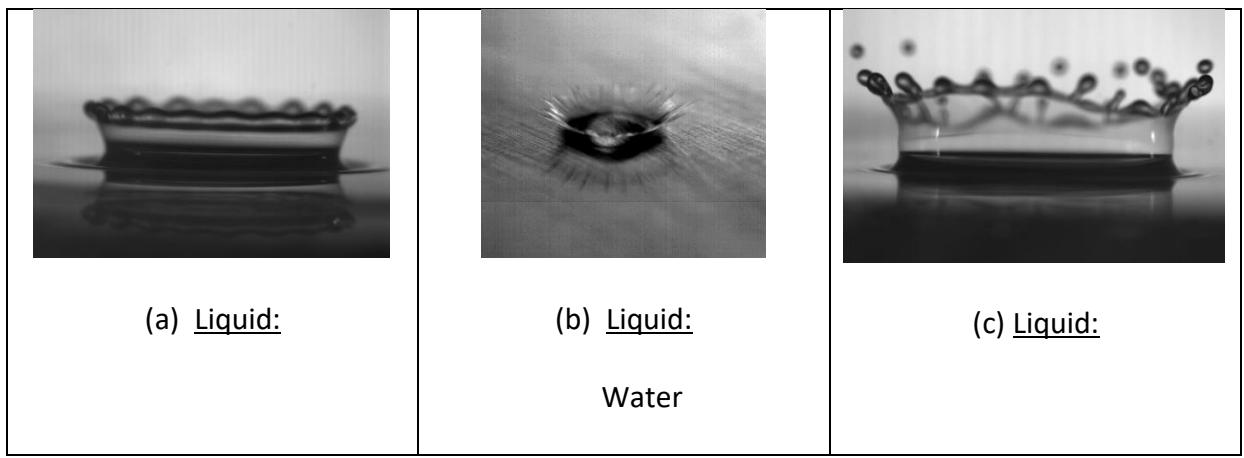

(a) Liquid:  (b) Liquid: Water  (c) Liquid:

| 50% by volume Propylene Glycol | | 50% by volume Propylene Glycol |
|---|---|---|
| Re = 1899 | Re = 8055.4 | Re = 2070.89 |
| We = 458.08 | We = 277.16 | We = 493.97 |

*Figure 1. Images showing (a) deposition (no splash), (b) prompt splash and (c) crown (delayed) splash along with respective properties*

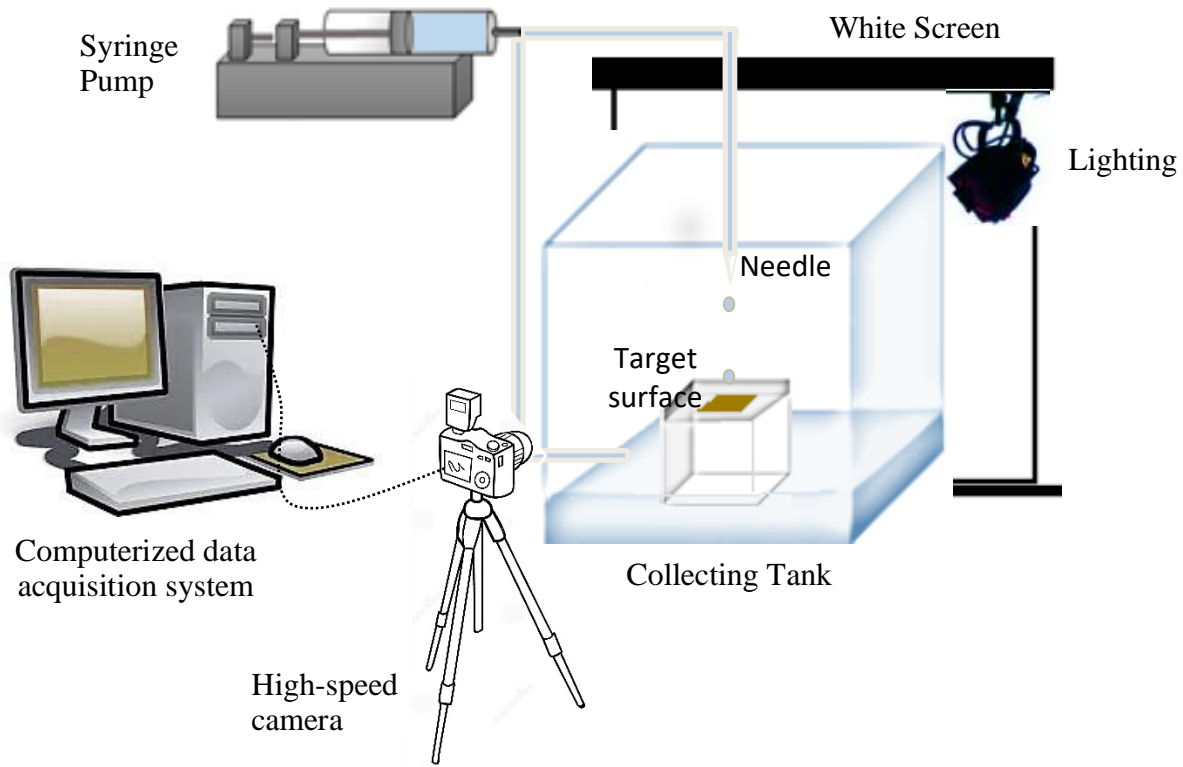

*Figure 2. Sketch of the experimental set-up*

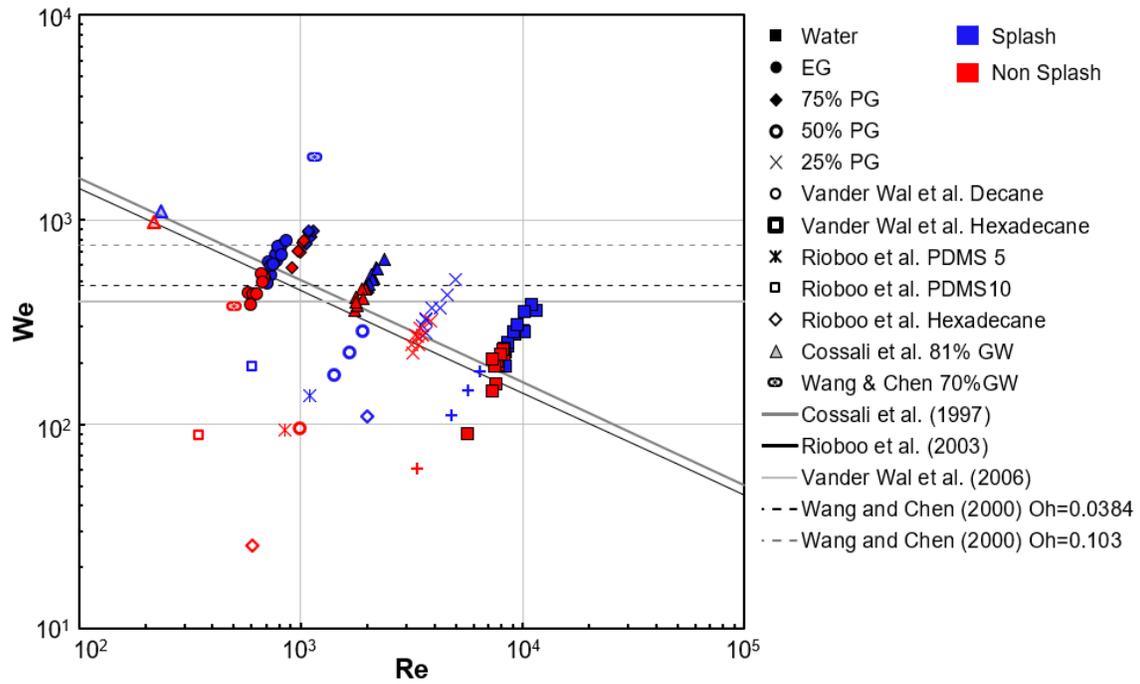

*Figure 3. Comparing past correlations with experimental data and with data from other studies for H* ≤ 0.1*

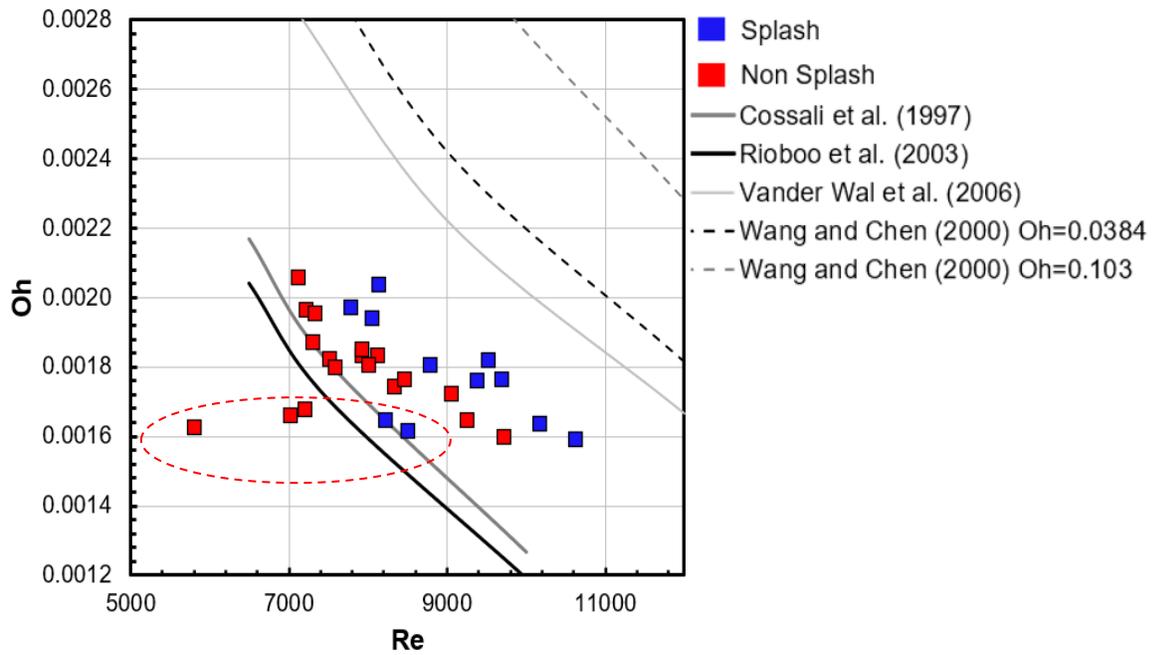

(a)

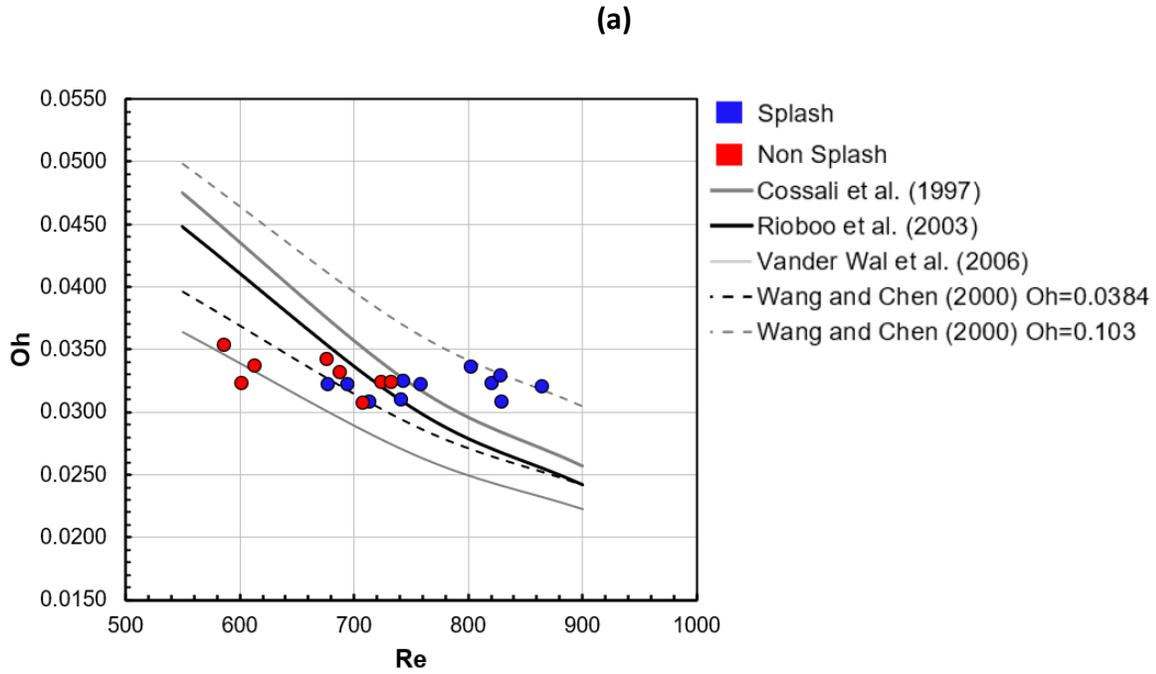

(b)

Figure 4. Comparing current experimental data with past correlation for (a) Water and (b) Ethylene glycol

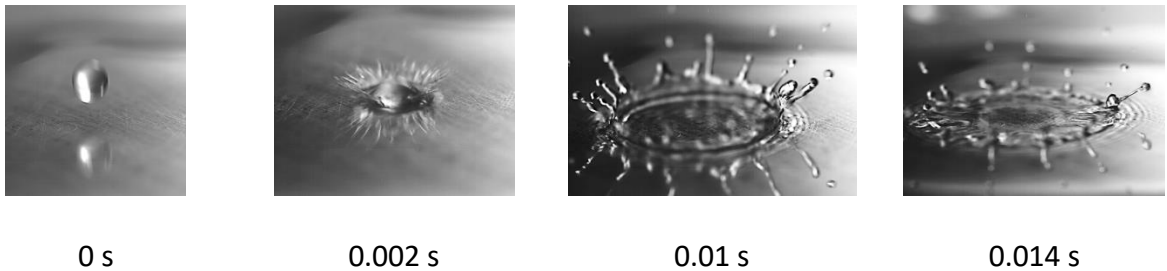

0 s      0.002 s      0.01 s      0.014 s

(a) Time progression of water drop impact on a thin film. d 4.6 mm v 2.2 m/s

Re = 10100 We = 305

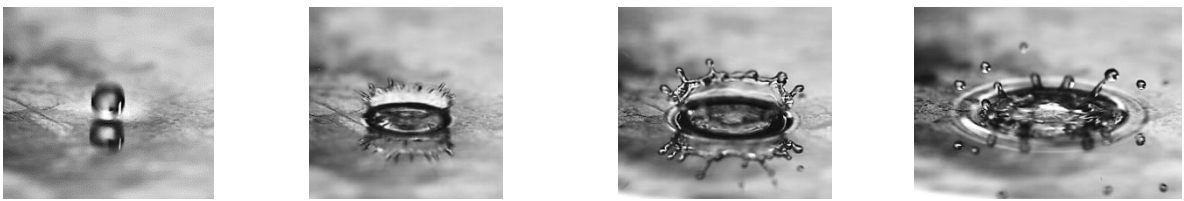

| 0 s | 0.002 s | 0.006 s | 0.012 s |

(b) Time progression of water drop impact on a thin film. d 3.6 mm v 2.5 m/s

Re = 8982 We = 308

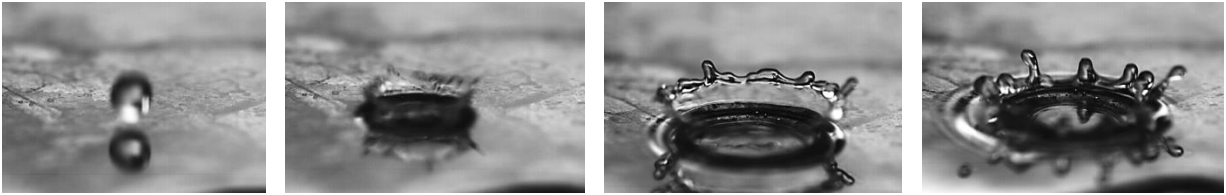

| 0 s | 0.002 s | 0.008 s | 0.0012 s |

(c) Time progression of water drop impact on a thin film. d 3.6 mm v 2.2 m/s

Re =7904 We = 239

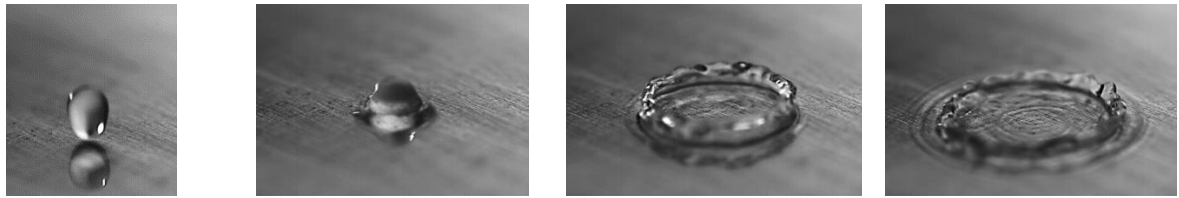

| 0 | 0.002 s | 0.01 s | 0.014 s |

(d) Time progression of water drop impact on a thin film. d 5.2 mm v 1.1 m/s

Re = 5709 We = 86

*Figure 5. Time Progression of drops of water*

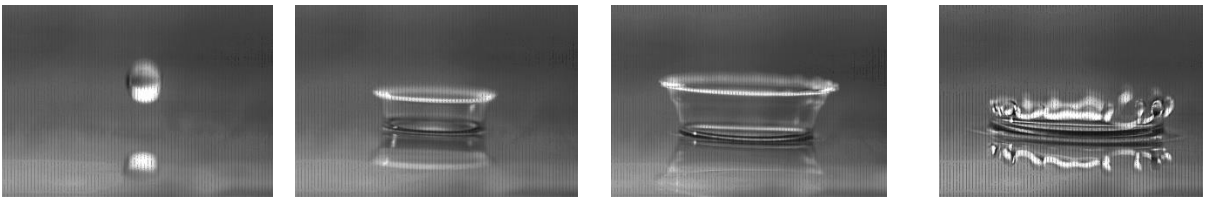

| 0 s | 0.004 s | 0.01 s | 0.018 s |

(a) Time progression of EG drop impact on a thin film. d 4.54 mm v 2.17 m/s

Re = 681.2 We = 491.7

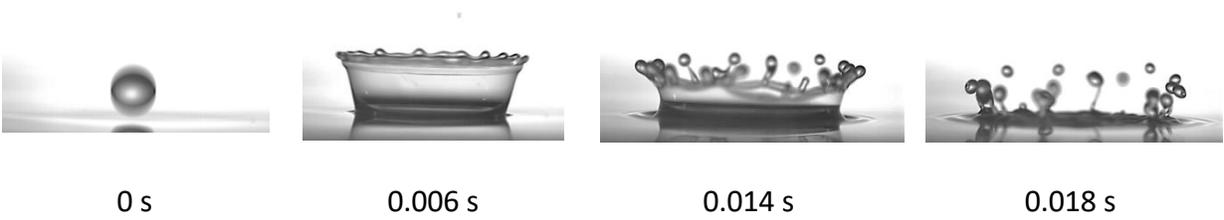

0 s　　　　　　　0.006 s　　　　　　　0.014 s　　　　　　　0.018 s

(b) Time progression of 50% PG drop impact on a thin film. d 4.4 mm v 2.18 m/s

Re = 1951 We = 470.5

*Figure 6. A comparison of impact behavior of EG and 50% PG drops*

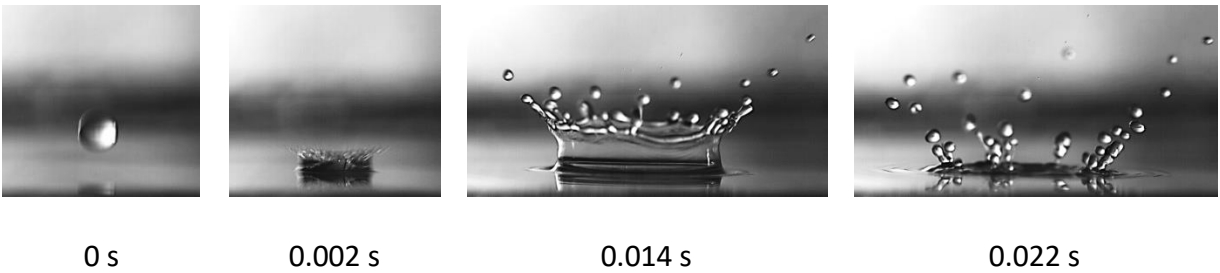

0 s　　　　　　　0.002 s　　　　　　　0.014 s　　　　　　　0.022 s

(a) Time progression of 25% PG drop impact on a thin film. d 4.5 mm v 2.5 m/s

Re = 5051.2 We = 525.2

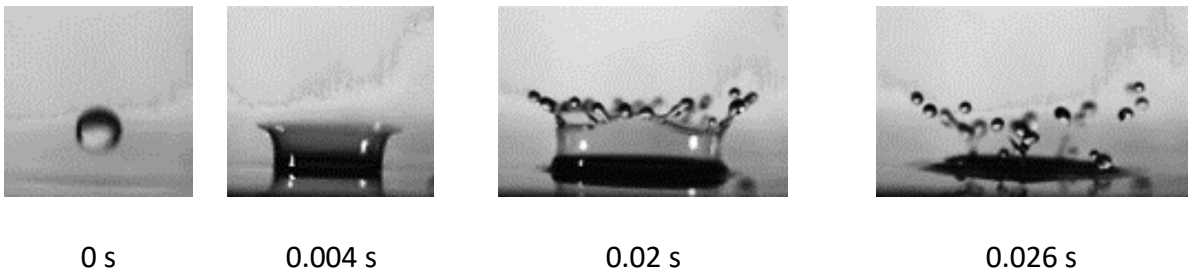

0 s　　　　　　　0.004 s　　　　　　　0.02 s　　　　　　　0.026 s

(a) Time progression of 75% PG drop impact on a thin film. d 5 mm v 2.45 m/s

Re = 1048 We = 749.6

*Figure 7. A comparison of impact behavior of 25% PG and 75% PG drops*

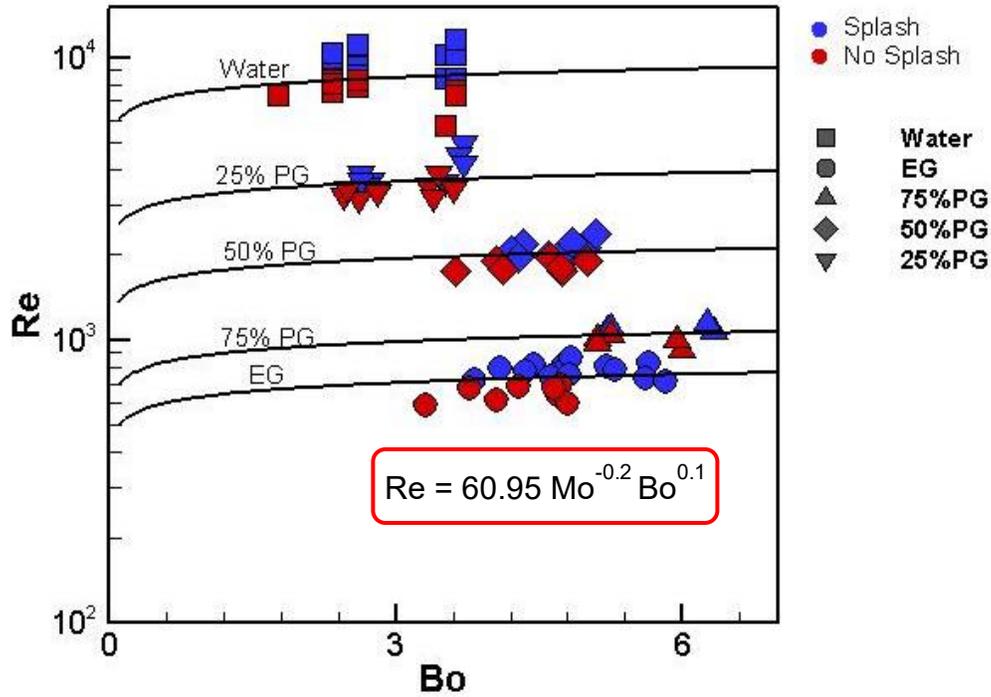

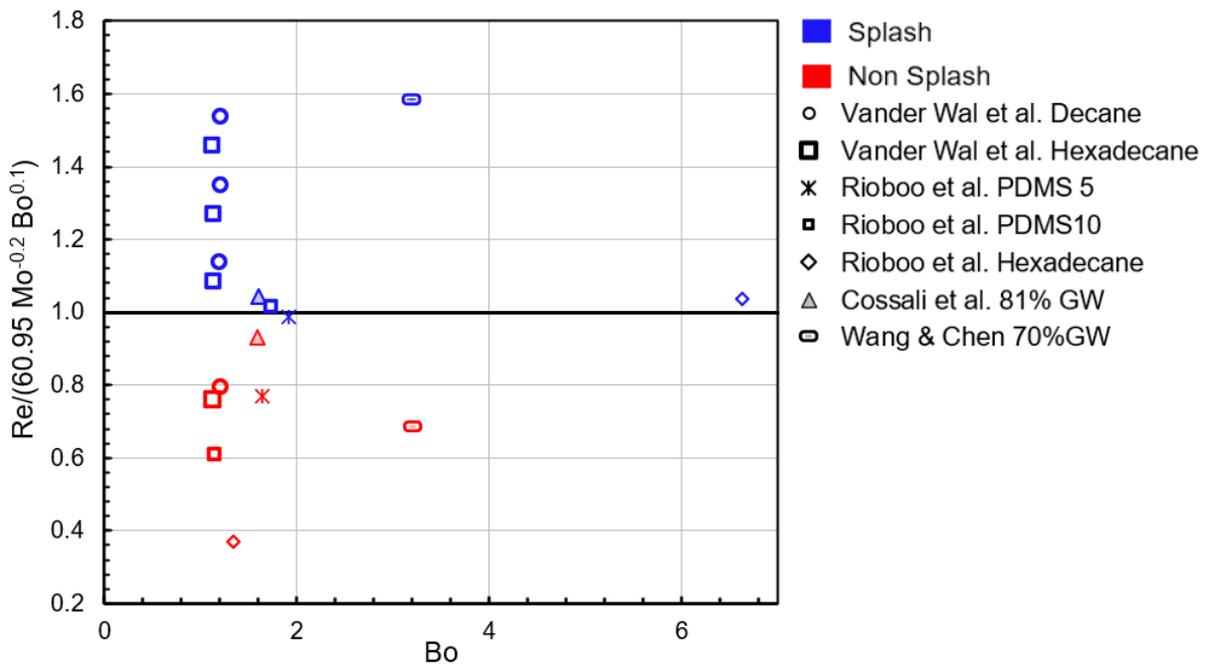

*Figure 8. Correlation (equation 2) performance with (a) current experimental data and (b) data from past literature*

*Table 1: Properties of the liquids used in the experiment*

| Liquids | Density (kg/m³) | Viscosity (Pa.s) | Surface Tension (N/m) |
|---|---|---|---|
| Water | 998.0 | 0.001 | 0.0728 |
| 25% by volume of Propylene Glycol (25% PG) | 1007.5 | 0.00255 | 0.0541 |
| 50% by volume of Propylene Glycol (50% PG) | 1017.0 | 0.005 | 0.0452 |
| 75% by volume of Propylene Glycol (75% PG) | 1026.5 | 0.012 | 0.0411 |
| Ethylene Glycol (EG) | 1113.2 | 0.0161 | 0.0484 |